\documentclass{elsart}
\usepackage{epsfig}

\begin{document}

\begin{frontmatter}

\title{Four-body DWBA calculations  of the Coulomb breakup of $^6$He}

\author[a]{R. Chatterjee},
\author[b]{P. Banerjee},
\author[a,c]{R. Shyam}

\address[a]{Theory Group, Saha Institute of Nuclear Physics,
1/AF Bidhan Nagar, \\
Calcutta - 700 064, INDIA.}
\address[b]{Physics Group, Variable Energy Cyclotron Centre,
1/AF Bidhan Nagar, \\
Calcutta - 700 064, INDIA.}
\address[c]{National Superconducting Cyclotron Laboratory, Michigan
State University, East Lansing, MI 48824, U.S.A.}

\begin{abstract}
We investigate the Coulomb breakup of the two-neutron halo nucleus
$^6$He by extending the framework of the finite range post form
distorted wave Born approximation theory introduced earlier for the  
description of the breakup of one-neutron halo nuclei. The model can use
the realistic three-body wave function for the $^6$He ground state.
Calculations have been performed for energy, total and parallel momentum
distributions of $^4$He fragment. The two-body $^4$He-neutron and
neutron-neutron correlations have also been calculated. 
Our results are compared with those of an adiabatic model of the Coulomb
breakup reactions. The two theories lead to almost identical results.
Comparisons are also made with the available experimental data. 
It is found that the pure Coulomb breakup can describe the
bulk of the data taken at below grazing angles. However, a rigorous
description of all the available data requires the consideration of
the nuclear breakup effects. 
\\
{\noindent 
{PACS numbers: 21.60.-n, 24.50.+g, 25.60.-t, 25.60.Gc}} 
   
\end{abstract}

\begin{keyword} {Coulomb breakup of $^6$He, two-neutron halo nucleus,
post form DWBA with finite range effects. } 
\end{keyword}

\end{frontmatter}
\renewcommand{\vec}[1]{\mbox{\boldmath$#1$\unboldmath}}
\newcommand{\cp}{\chi^{(+)}}
\newcommand{\cm}{\chi^{(-)*}}
\newcommand{\cmm}{\chi^{(-)}}
\newcommand{\ri}{{\bf r}_i}
\newcommand{\ak}{{\bf k}_a}
\newcommand{\bq}{{\bf k}_b}
\newcommand{\lB}{\hat{l}\beta_{lm}}
\newcommand{\B}{\beta_{lm}}
\newcommand{\rc}{{\bf r}_c}
\newcommand{\rb}{{\bf r}_b}
\newcommand{\cq}{{\bf k}_c}
\newcommand{\bm}{\bibitem}
\newcommand{\we}{\Psi^{(+)}_a(\xi_a,{\bf R},{\bf r},{\bf r}_i)}
\newcommand{\fa}{ _1F_1(-i\eta_a,1,i(k_ar_i - {\bf k}_a \cdot {\bf r}_i))} 
\newcommand{\fB}{ _1F_1(-i\eta_b,1,i(k_br_i + {\bf k}_b \cdot {\bf r}_i))}
\newcommand{\faa}{ _1F_1(-i\eta_a,1,i(k_ar_b -{\bf k}_a \cdot {\bf r}_b))} 
\newcommand{\fBB}{ _1F_1(-i\eta_b,1,i(k_br_b +{\bf k}_b \cdot {\bf r}_b))}  
\newcommand{\yr}{Y^l_m(\hat{\bf r})}
\newcommand{\ytr}{Y^{\lambda}_\mu(\hat{\bf R})}
\newcommand{\ytk}{Y^l_m(\hat{\bf k})}
\newcommand{\yko}{Y^{\lambda}_\mu(\hat{\bf k}_1)}

\section{Introduction} 
Several nuclei near the drip lines have been found to have properties
which are strikingly different from those of their stable counterparts. 
These nuclei have a halo structure in their ground
states in which loosely bound valence nucleon(s) has (have) a large
spatial extension with respect to the respective core
\cite{tani96,hans95,zhuk93,hans87,tani85}. The two-neutron halo nuclei,
of which $^6$He is a prominent example, have attracted a lot of attention.
These are Borromean three-body systems, where all the two-body
subsystems are unbound \cite{zhuk93,fedo94}. Clearly, the interaction 
of valence neutrons is vital for the stability of the two-neutron halo
nuclei. They provide interesting examples of the three-body problem 
in the limit of very weak binding \cite{eff90}, and the three-body 
methods (within both Faddeev and cluster type approaches) have been
extensively applied to study them 
\cite{zhuk93,leh83,suz91,bang92,dani93,cso93,fun94,dani94,aoy95,dani97,dani98,ers97,ian98}.  
Other approaches like non-relativistic \cite{gome92}
and relativistic \cite{koe91,zhu94} mean field models and many-body calculations
\cite{pan97} have also been used to investigate the structure of such nuclei. 

Breakup reactions, in which the valence neutrons are removed from the
core in the interaction of halo nuclei with stable targets,
provide detailed information on their structures \cite{raja00}.
The study of the breakup reaction of $^6$He is interesting due to several
reasons. It leads to the excitation of the continuum of this 
nucleus, which has received much attention recently due to the experimental
discovery \cite{naka00} of the soft dipole resonance (SDR), signifying
the dipole oscillation between the core cluster and the valence neutrons. 
Although the SDR was predicted long ago \cite{hans87,suz91},
properties of this resonance measured in Ref. \cite{naka00} are at
variance with those predicted by the detailed three-body calculations
\cite{dani97}. The breakup reaction of this nucleus gives access to the study
of the neutron-neutron ($n$-$n$) and neutron-alpha particle ($n$-$^4$He)
correlations, which are
important because they provide a critical test of the
dineutron model, in which the halo neutrons were treated as a point like 
``dineutron" cluster orbiting the core. This model has been used in
earlier theoretical studies, both semiclassical \cite{cant93}
and quantum mechanical \cite{ban92,bert91}, of the breakup of the
two-neutron halo nuclei. Furthermore, the study of the neutron-$^4$He
correlation is expected to constrain the neutron-$^4$He interaction used
in the three-body calculations of the $^6$He ground state. The study of the
total and longitudinal momentum distributions of the charged fragment is
expected to yield information on the matter radii of the halo.
The breakup of $^6$He is also important from the viewpoint of astrophysics
as it can be used \cite{baur00} to extract the rate of the
$^4$He($2n,\gamma)^6$He($2n,\gamma)^8$He reactions
which could be a possible route for bridging the gap at the mass number,
$A$ = 5 and 8.   

Several calculations of the nuclear and Coulomb breakup 
\cite{gar97a,gar98,gar00a,gar00b} of $^6$He, based on a model
which uses a sudden approximation \cite{gar97b,gar99}, have been reported
recently. Calculations of the nuclear breakup of $^6$He within the 
approximate eikonal model is reported in Ref. \cite{ber98}. Due to the
nature of the approximations involved in these models, their applicability
is limited to reactions at higher beam energies ($>$ 200 MeV/nucleon). 
Moreover, they have been applied to describe only a selected set of 
observables. In the distorted wave Born approximation
(DWBA) calculations of the breakup of $^6$He reported in \cite{ers00},
the interaction appearing in the kernel of the prior DWBA T-matrix 
has been evaluated in  the impulse approximation which limits its
applicability to higher beam energies.
 
In this paper, we calculate the Coulomb breakup of $^6$He treating the 
final state as a four-body system. Our formalism uses   
the framework of the post form distorted wave Born approximation 
which is an extension of the theory developed earlier \cite{raja00}
to describe the breakup of one-neutron halo nuclei.  
Finite range of the interaction between projectile constituents 
is treated by making an effective local momentum approximation (ELMA) 
on the distorted wave of the charged core in the final
channel.  It was shown in Ref.~\cite{raja00} that this
approximation is well satisfied for the breakup of one-neutron halo
nuclei. Furthermore, this gave results \cite{raja00} which were similar
to those obtained within an adiabatic model (AD) \cite{toste98} of
Coulomb breakup reactions. This theory allows the use of a realistic
three-body wave function for the projectile ground state and 
can be applied to breakup reactions with beam energies ranging from
below the Coulomb barrier to higher values up to which the non-relativistic
treatment is valid.  We have used our theory to analyze almost all
types of the breakup observables measured in the reaction of $^6$He
on heavy target nuclei. We compare our results 
with those of the AD and dineutron models.

This paper is organized in the following way. Our formalism is presented
in the next section. The structure models of  
$^6$He are discussed in section 3. The results of our calculations and
discussions  are presented in section 4. A summary and the conclusions
of our work are given in section 5.
\section{Formalism}
We consider the breakup reaction 
$ a + t \rightarrow b + n_1 + n_2 + t $, where the 
projectile $a$ breaks up into fragment $b$ (the charged core) and the two 
valence neutrons ($n_1$ and $n_2$) in the Coulomb field of the 
heavy target $t$. We consider only the  elastic breakup mode in which 
the target nucleus remains in its ground state during the breakup process.
\begin{figure}[ht]
\begin{center}
\mbox{\epsfig{file=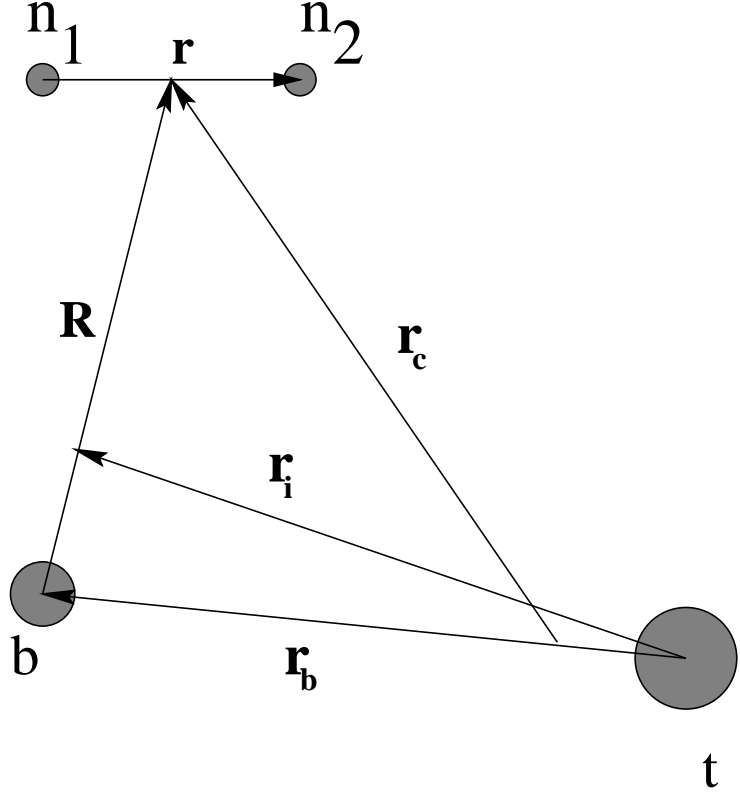,width=.70\textwidth}}
\end{center}
\caption{
The four -- body coordinate system. $n_1,n_2$ are the valence neutrons. $b$
and $t$ represent the charged core and the target respectively.}
\label{fig:fig1}
\end{figure}
The system of coordinates is shown in Fig. 1. The position vectors 
satisfy the relations
\begin{eqnarray}
\rb &=& \ri - \alpha{\bf R},~~~~ \alpha = {2m_n\over {2m_n+m_b}}, \\ 
\rc &=& \beta\ri +\delta{\bf R},~~~~ \beta = {m_t\over {m_b+m_t}}, 
~~~\delta = (1 - \alpha\beta).
\end{eqnarray}

The post form $T$ - matrix for this case is given by
\begin{eqnarray}
T & = &\langle
\cmm_{b-t}(\bq,{\bf r}_b)\phi_b(\xi_b)
\cmm_{c-bt}(\cq,\rc)\phi_{n_1n_2}({\bf k},{\bf r},\xi_{{n_1n_2}})
|V({\bf R}, {\bf r})|\nonumber \\
&\times & \we \rangle,  
\end{eqnarray}
where $c$ represents the c.m. of the $n_1-n_2$ system.
$\cmm_{b-t}$ is the Coulomb distorted wave describing 
the motion of the charged core ($b$)
with respect to target $t$ with incoming wave boundary condition 
and $\cmm_{c-bt}$
describes the c.m. motion of the valence neutrons with respect to the 
c.m. of the $b$-$t$ system. $\phi$'s
are the internal state wave functions of the concerned particles with
$\xi$'s being the internal coordinates. $\bq$ and $\cq$ are the Jacobi wave
vectors
conjugate to $\rb$ and $\rc$ respectively.  $\we$ represents the
exact four-body scattering wave function of the projectile
with an outgoing wave boundary condition.  $\ak$ is the  
wave vector associated with it. 
$V({\bf R},{\bf r})$ is the sum of the
two-body interactions between the projectile constituents, given by 
\begin{eqnarray}
V({\bf R},{\bf r}) = V_{n_1n_2}({\bf r}) + V_{bn_1}({\bf R}-{\bf r}/2) +
V_{bn_2}({\bf R}+{\bf r}/2).
\end{eqnarray}
\subsection{Transition amplitude within the finite range
distorted wave Born approximation (FRDWBA)}
In DWBA theory the breakup states are assumed to be weakly coupled
and hence this coupling is treated only up to the first order. Thus 
we can write 
\begin{eqnarray}
&&\we = \phi_a(\xi_a, {\bf R}, {\bf r})\cp_{a-t}(\ak,\ri),
\end{eqnarray}  
where $\phi_a(\xi_a, {\bf R}, {\bf r})$ represents the (three-body)
bound state wave function of the
projectile and $\cp_{a-t}(\ak,\ri)$ is the Coulomb distorted scattering 
wave describing the relative motion of the c.m. of the projectile with
respect to the target with outgoing wave boundary condition.
Substituting Eq. (5) into Eq. (3) and writing the latter in the 
integral form we get,  
\begin{eqnarray}
T^{DWBA}&=&\int\int\int\int d\xi d{\bf R} d{\bf r}
d\ri \cm_{b-t}(\bq,{\bf r}_b)\phi^*_b(\xi_b) \cm_{c-bt}(\cq,\rc)
\nonumber \\ &\times &
\phi^*_{n_1n_2}({\bf k},{\bf r},\xi_{{n_1n_2}})
V({\bf R},{\bf r})\phi_a(\xi_a, {\bf R}, {\bf r})\cp_{a-t}(\ak,\ri).   
\end{eqnarray}
We introduce the finite range effects in DWBA by an effective local 
momentum approximation by writing the distorted wave in the final channel
as \cite{raja00}
\begin{eqnarray}
\cm_b(\bq,{\bf r}_b) & = & e^{i\alpha{\bf K} \cdot {\bf R}}\cp_b(-{\bq},\ri),
\end{eqnarray} 
where ${{\bf K}}$ is an effective local momentum whose direction is
taken to be the same as that of the asymptotic momentum ${\bf k}_b$,
since the calculated cross sections are almost independent 
of this quantity, as is shown in Appendix A.
The magnitude of ${\bf {K}}$ is given by
\begin{eqnarray} 
{ {K}}(R_d) &=& {\sqrt {{2\mu_{b-t}\over \hbar^2}(E - V(R_d))}}.
\end{eqnarray}
In Eq. (8) $\mu_{b-t}$ is the reduced mass of the $b$-$t$ system,
$E$ is the energy of particle $b$ relative to the target in the
c.m. system and $V(R_d)$ is the  Coulomb potential
between $b$ and the target at a distance $R_d$. Thus, the magnitude
of ${\bf K}$ is evaluated at a separation $R_d$ (taken to be 10 fm
in all the numerical calculations shown in the 
next section), which is held fixed for all the values of $r$.   
We refer to Appendix A for further discussions of the validity
of this approximation for the reactions studied in this paper.

In Eq. (6), the Coulomb distorted waves are given by
\begin{eqnarray}
\cp_{a-t}(\ak,\ri)& = & e^{-\pi\eta_a/2}\Gamma(1+i\eta_a)
e^{i\ak \cdot \ri}{\fa},\\ 
\cm_{b-t}(\bq,{\bf r}_b)& = & e^{i\alpha{\bf K} \cdot {\bf R}}
e^{-\pi\eta_b/2}\Gamma(1+i\eta_b)e^{-i\bq \cdot \ri}
\nonumber \\ &\times &
{\fB}. 
\end{eqnarray}
For pure Coulomb breakup, the interactions between two uncharged valence
particles as well as between their c.m. and that of the $b$-$t$ system are
zero. Hence we take plane waves for the $c$-$bt$ and $n_1$-$n_2$ relative
motions:
\begin{eqnarray}
\cm_{c-bt}(\cq,{\bf r}_c) 
         & = & e^{-i\beta\cq \cdot \ri}e^{-i\delta\cq \cdot {\bf R}},\\
\phi^*_{n_1n_2}({\bf k},{\bf r},\xi_{12}) & = & e^{-i{\bf k} \cdot {\bf r}}
\phi^*_{n_1n_2}(\xi_{12}).
\end{eqnarray}
Substituting Eqs. (9-12) into Eq. (6) we get,
\begin{eqnarray}
T^{FRDWBA}&=& e^{-\pi(\eta_a+\eta_b)/2}\Gamma(1+i\eta_a)\Gamma(1+i\eta_b)
\nonumber\\ 
&\times [&\int d\ri e^{i(\ak-\bq-\beta\cq) \cdot \ri} {\fa}\nonumber \\
&\times &\fB] \nonumber\\ 
&\times [&\int d{\bf r} d{\bf R} d\xi e^{-i(\delta\cq - \alpha{
{\bf K}}) \cdot {\bf R}}
e^{-i{\bf k} \cdot {\bf r}}\phi^*_b(\xi_b)\phi^*_{n_1n_2}(\xi_{12})
V({\bf R},{\bf r})
\nonumber \\ &\times &\phi_a(\xi_a, {\bf R}, {\bf r})]. 
\end{eqnarray}
In Eq. (13), the first integral is associated with the dynamics of the
reaction which can be expressed analytically in terms of the
Bremsstrahlung integral \cite{nord}. The second one (to be expressed as 
$I_f^{FRDWBA}$) contains the information about the structure of 
the projectile and is known as the form factor.
Defining $\bf {\delta\cq - \alpha{\bf K}}$ = ${\bf {k_1}}$, we can 
write
\begin{eqnarray}
I_f^{FRDWBA}&=& \int d{\bf r} d{\bf R}
e^{-i{\bf {k_1}} \cdot {\bf R}}e^{-i{\bf k} \cdot {\bf r}}
[\int d\xi\phi^*_b(\xi_b)\phi^*_{n_1n_2}(\xi_{12})
V({\bf R},{\bf r})\nonumber \\ 
&\times &\phi_a(\xi_a, {\bf R}, {\bf r})]. 
\end{eqnarray}
The function $\Phi_a(\xi_a, {\bf R}, {\bf r}) = V({\bf R},{\bf r})\phi_a(\xi_a,
{\bf R}, {\bf r})$ has the following
representation in configuration space \cite{ban98,jim95}:
\begin{eqnarray}
\sum_{l\lambda LS}\Phi_{l\lambda
LS}(R,r)[[\yr\otimes\ytr]_L\otimes\chi_S]_{JM}.
\end{eqnarray} 
In Eq. (15), the relative orbital angular momentum $\lambda$
between the core and c.m. of the two neutrons
is coupled with the relative orbital angular momentum $l$ between
the neutrons to give $L$, which,
in turn, is coupled with the total intrinsic spin $S$ of the two valence
neutrons to give the resultant total angular momentum $J$ of the projectile
with projection $M$. For $^6$He, $J=M=0$. 

Hence
\begin{eqnarray}
&&\int d{\bf r} d{\bf R} e^{-i(\delta\cq - \alpha{{\bf K}}) \cdot {\bf R}}
e^{-i{\bf k} \cdot {\bf r}}
V({\bf R},{\bf r})\phi_a(\xi_a, {\bf R}, {\bf r})
\\ \nonumber
&=& 4\pi\sum_{l\lambda LS}(-i)^{l+\lambda}
\Phi_{l\lambda LS}(k,k_1)
[[\ytk\otimes\yko]_L\otimes\chi_S]_{00}, 
\end{eqnarray}
where
\begin{equation}
\Phi_{l\lambda LS}(k,k_1) = 4\pi\int r^2dr\int R^2dR j_l(kr)
j_{\lambda}(k_1R)\Phi_{l\lambda LS}(R,r).
\end{equation}

The modulus square of $I_f^{FRDWBA}$ is given by 
\begin{eqnarray}
&&|I_f^{FRDWBA}|^2 = ({4\pi})^2 \sum_{ll'\lambda\lambda' L \Lambda}
(-)^{L+\Lambda}\Phi_{l\lambda LL}(k,k_1)\Phi_{l'\lambda' LL}(k,k_1)
P_{\Lambda}(\cos\Theta)
\nonumber \\ && \times
{\hat l}{\hat l'}{\hat \lambda}{\hat \lambda'}{\hat \Lambda}^2 
W(l \lambda l' \lambda';L \Lambda)
\left(\begin{array}{ccc}l&l'&\Lambda\\
0&0&0\end{array}\right)
\left(\begin{array}{ccc}\lambda&\lambda'&\Lambda\\
0&0&0\end{array}\right),
\end{eqnarray}
where $\Theta$ is the angle between vectors ${\bf k}_1$ and
${\bf k}$, and ${\hat n} = \sqrt{2n + 1}$.

The dominant angular momentum configurations of $^6$He contributing
to Eq. (18) are  $\Phi_{0000}(k,k_1)$ and $\Phi_{1111}(k,k_1)$ 
(with probabilities of $\approx 0.84$ and $\approx 0.11$, 
respectively). Thus we have \cite{ban98}
\begin{eqnarray}
&&|I_f^{FRDWBA}|^2 = \Phi^2_{0000}(k,k_1) + \Phi^2_{1111}(k,k_1)
- \Phi^2_{1111}(k,k_1)P_2(\cos\Theta).
\end{eqnarray}
The differential cross section for the breakup process is given by
\begin{eqnarray}
{{d^5\sigma}\over {dE_b d\Omega_b dE_{n_1} d\Omega_{n_1} d\Omega_{n_2}}}&=&
{{2\pi}\over {\hbar v_a}}\rho(E_b,\Omega_b,E_{n_1},\Omega_{n_1},\Omega_{n_2})
\nonumber \\ &\times &|T^{FRDWBA}|^2,
\end{eqnarray}
where $v_a$
is the relative velocity of the projectile in the initial channel.

The four-body phase space factor is given by 
(see Appendix B) 
\begin{eqnarray}
\rho(E_b,\Omega_b,E_{n_1},\Omega_{n_1},\Omega_{n_2}) = 
{h^{-9}m_tm_bm_{n_1}m_{n_2}p_bp_{n_1}p_{n_2}\over m_{n_2} +
m_t -m_{n_2}{{\bf p}_{n_2} \cdot ({\bf p}_{tot}
 -{\bf p}_b - {\bf p}_{n_1})\over p_{n_2}^2}},
\end{eqnarray}
In Eq.(21), ${\bf p}_i$'s ($i = t, b, n_1, n_2$) are
the momenta appropriate to the frame of interest
(i.e. laboratory or c.m.) and
${\bf p}_{tot} = {\bf p}_{t} + {\bf p}_{b} + {\bf p}_{n_1} + {\bf p}_{n_2}$.

\subsection{Transition amplitude within the adiabatic (AD) model}

The adiabatic model of Coulomb breakup assumes that the
important excitations of the projectile nucleus are in the low energy
continuum, i.e. the breakup configurations excited by the projectile-target
interaction are low energy $b$+$n_1$+$n_2$ relative motion states. This
allows the four-body wave function $\we$ to be written as \cite{ban98}
\begin{eqnarray}
\we = \phi_a(\xi_a, {\bf R}, {\bf r})e^{i\alpha \ak \cdot {\bf R}}
        \cp_{a-t}(\ak,\rb).
\end{eqnarray}  
The AD model is a non-perturbative theory and it does not make the weak 
coupling approximation of DWBA. However, the initial and final state
Coulomb interactions are included to all orders in both the formalisms.
It may be noted that while the AD model necessarily requires the
valence nucleons to be chargeless, the FRDWBA can, in principle,
be applied to cases where both the core and the valence particles are
charged.

The expression for the transition amplitude for the Coulomb breakup 
of a Borromean nucleus within the adiabatic model is given by \cite{ban98}
\begin{eqnarray}
T^{AD}&=& e^{-\pi(\eta_a+\eta_b)/2}\Gamma(1+i\eta_a)\Gamma(1+i\eta_b)
 \nonumber \\
&\times &[\int d\ri e^{i(\ak-\bq-\beta\cq) \cdot \rb} {\faa} \nonumber \\
&\times & \fBB] \nonumber \\ &\times & 
[\int d{\bf r} d{\bf R} d\xi e^{-i(\cq - \alpha{ {\bf k}_a}) \cdot {\bf R}}
e^{-i{\bf k} \cdot {\bf r}}\phi^*_b(\xi_b)\phi^*_{n_1n_2}(\xi_{12})
V({\bf R},{\bf r}) \nonumber \\ &\times &
\phi_a(\xi_a, {\bf R}, {\bf r})]. 
\end{eqnarray}
It is evident that except for the form factor, the transition amplitudes
within the FRDWBA and AD formalisms are the same. The form factor 
in the AD case is given by
\begin{eqnarray}
I_f^{AD}& = &\int d{\bf r} d{\bf R}e^{-i(\cq - \alpha{
{\bf k}_a}) \cdot {\bf R}} 
e^{-i{\bf k} \cdot {\bf r}}[\int d\xi\phi^*_b(\xi_b)\phi^*_{n_1n_2}(\xi_{12})
V({\bf R},{\bf r})\nonumber \\ &\times &
\phi_a(\xi_a, {\bf R}, {\bf r})].
\end{eqnarray} 
It may be noted that Eq. (23) can also be obtained within the DWBA model
by making ELMA to the Coulomb distorted wave in the initial channel of the
reaction, and by evaluating the local momentum at $R \ = \ \infty$ with 
its direction being the same as that of the projectile.
 
\section{\bf Structure models}

The three-body wave function for the ground state (g.s.) of $^6$He   
used in our calculations (which was provided to us by the authors of
Ref. \cite{dani98}) has been obtained \cite{dani98}
by solving the Faddeev equation with the hyperspherical
harmonic expansion method which included a maximum of 20 hyperharmonics.
In these calculations the $n$-${^4}He$  interaction was
taken from Ref.~\cite{bang79}, and for the $n$-$n$ interaction
the Gogny-Pires-Tourreil potential \cite{gog70} with spin-orbit and
tensor components was used. The two-neutron separation energy and
root mean square (rms) matter radius of $^6$He predicted by this model
were  0.975 MeV and 2.50 fm, respectively. The $\alpha$-particle rms
radius was taken to be 1.49 fm.

We would like to emphasize that in this paper our main intention is
to develop a post form FRDWBA theory of the Coulomb breakup of the
Borromean nuclei where a three-body wave function for their ground state
can be used. This enables the breakup observables to be employed as additional
constraints on the three-body structure models of these systems. Performing
structure calculations is outside the scope of this paper.
Nevertheless, it would be worthwhile to use the g.s. wave function of
$^6$He calculated by other authors (see eg. \cite{cob97} where the 
calculations for the $^6$He have been done within  
the Faddeev approach using different sets of interactions),
to see the sensitivity of our results on the structure models of $^6$He.  

For the dineutron model calculations, a point-like `dineutron'
was assumed to be bound to the $^4$He core in a 1$s$ state. The corresponding 
interaction is described by a Woods-Saxon potential
(with radius and diffuseness parameters of 1.15 fm and 0.5 fm respectively)
whose depth is adjusted to reproduce the binding energy of 0.975 MeV.
The resulting $^6$He wave function has the rms radius of 2.47 fm, with the
same $\alpha$-particle rms radius as that mentioned above.

\section{\bf Results and discussions}

\subsection{$\alpha$-particle energy distribution}

\begin{figure}[ht]
\begin{center}
\mbox{\epsfig{file=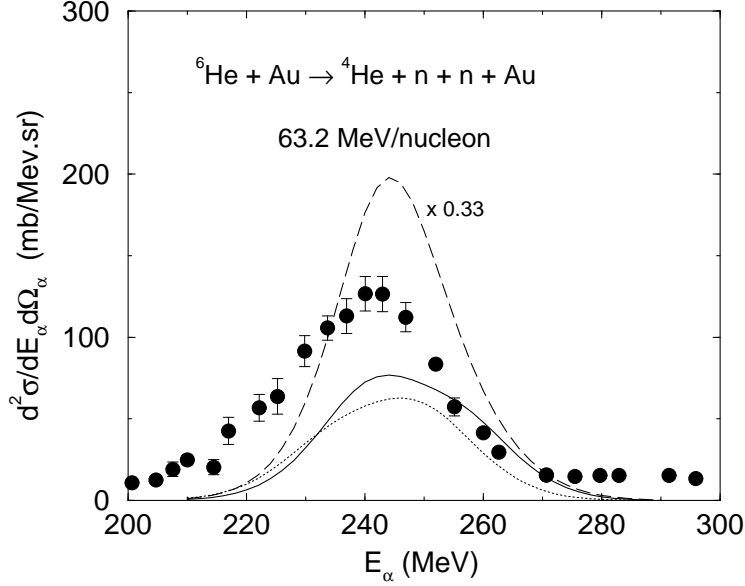,width=.70\textwidth}}
\end{center}
\caption{Calculated energy distributions of $\alpha$-particle in the elastic
breakup of $^6$He on a Au target at 63.2 MeV/nucleon at $\theta_{\alpha}$ = 5$^
\circ$. The solid, dotted and dashed curves represent the results
of the FRDWBA, AD and dineutron models respectively.
The results of the di-neutron model have been plotted after multiplying 
the actual cross sections by 0.33.
The data have been taken from \protect\cite{bala}.}
\label{fig:fig2}
\end{figure}

In Fig. 2, we show a comparison of the calculated and measured  
$\alpha$-particle energy spectra at $\theta _{\alpha}$ = 5$^{\circ}$
in the elastic breakup of $^6$He on a Au target at the beam energy of
63.2 MeV/nucleon. The data are taken from Ref.~\cite{bala}. These  
cross sections have been obtained by performing the 
neutron angular integrations (in Eq. (20)) up to 20$^\circ$ and 360$^\circ$
in the theta and phi planes, respectively. The results remained almost
unaffected by increasing the theta angle further.
All the kinematically allowed values were included in the neutron
energy integration. The solid, dotted and dashed lines show the
results of calculations performed within the FRDWBA, AD and dineutron
models respectively. The results of the last model are plotted after
multiplying the actual cross sections by the factor shown in the figure.
It may be noted that the theoretical spectrum in all the cases is 
shifted backwards by 10 MeV to compensate for the energy loss due to the
target thickness \cite{bala,ban98}. It is seen that the 
FRDWBA results are about 20$\%$ larger than those of the AD
model at the $\alpha$ particle energies near and above the peak region.  
However, both of them are lower than the experimental data.
The grazing angle for the reaction studied in this 
figure is about 3.5$^\circ$. Therefore, nuclear breakup 
effects are expected to make significant contribution to the cross
sections at the angle for which these data are taken. This could explain the
pure Coulomb breakup results being smaller than the experimental
cross sections, particularly in the peak region. 

It may be noted that the `dineutron' model overestimates the measured
cross sections by a factor of about 3 (the dashed line). This result
can be understood qualitatively by recalling that the magnitudes of
the calculated cross sections are dependent on the mean square
distance between the $\alpha$-particle and the center of mass of 
$^6$He \cite{sak}.
In case of the three-body wave function, this distance is 1.428 fm$^2$,
whereas for the `dineutron' model, it is 2.32 fm$^2$. The larger
cross section obtained in the latter model, therefore, may be attributed
to this difference.

Measurements performed below the grazing angle are expected to be less
affected by the nuclear breakup. One could then hope  
to test the significance of the small but noticeable difference
seen in the pure Coulomb breakup cross sections
calculated within the FRDWBA and AD models. However, it can  
already be seen that the peak position in the cross section calculated
within the former is somewhat shifted to smaller energies as compared
to that of the latter and is in a slightly better agreement 
with the peak position of the data.  
\subsection{Coulomb part of total two-neutron removal cross sections}
\begin{figure}[ht]
\begin{center}
\mbox{\epsfig{file=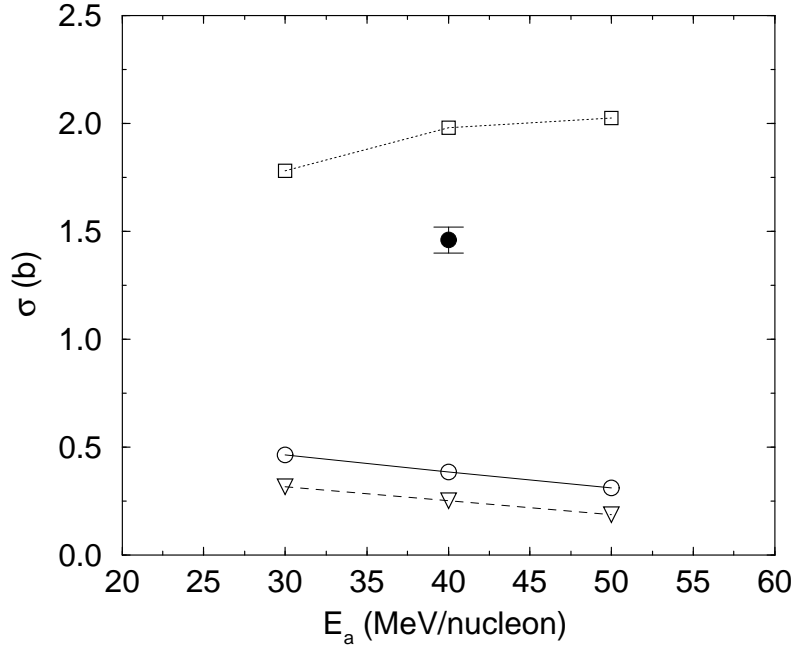,width=.75\textwidth}}
\end{center}
\caption{Calculated Coulomb part of total two-neutron removal cross sections
for breakup of $^6$He on a Pb target at three different beam energies of 30,
40 and 50 MeV/nucleon. The open triangles, open circles and open squares 
show the results of the AD, FRDWBA and dineutron models respectively.
The data point is the average of the experimental total two-neutron
removal cross sections measured between the beam energies
of 28 MeV/nucleon to 52 MeV/nucleon (\protect\cite{warn}).}
\label{fig:fig3}
\end{figure}

In Fig. 3, we show the results of calculations for the Coulomb part
of the total two-neutron removal cross section
at three different beam energies of 30, 40 and 50 MeV/nucleon for the
breakup reaction of $^6$He on a $^{208}$Pb target. 
The open circles, open triangles and open squares show the results
of the FRDWBA, AD and dineutron models respectively. With increasing
beam energy the cross sections of first two models decrease, which is
consistent with the earlier observations (see {$\it e.g.$} \cite{ban92}).
We note that here also the results of these two models differ
from each other by 20-25$\%$, but both of them are smaller
than the rough estimates for the pure Coulomb breakup cross sections given
in Ref.~\cite{warn}. However, such estimates are unlikely to be accurate 
for nuclei having a halo structure as discussed in \cite{ban99,dasso98}.
 
That the nuclear breakup contributions are substantial in the case of
breakup of $^6$He even on a heavy target, is evident from this figure,
where we also show (solid circle) the experimental value of the measured
total two-neutron removal cross section($\sigma_{-2n}^{exp}$),
which is the average of the measurements performed at beam energies
between 28 to 52 MeV/nucleon. Our theoretical values correspond only to
the Coulomb part of the elastic breakup cross section, while 
$\sigma_{-2n}^{exp}$ has contributions from
the Coulomb, nuclear as well as their interference terms. Moreover, the
nuclear breakup consists also of the inelastic breakup mode \cite{shyam98} 
(which is called as the stripping process in Ref. \cite{ber98}). 
Therefore, to understand the data of Ref. \cite{warn}, it is essential
to calculate the nuclear breakup cross sections on the same footing as
the Coulomb one. Similarly, in the breakup of $^6$He on the Pb target at
the higher beam energy of 240 MeV/nucleon, the calculated pure Coulomb
two-neutron removal cross section is 62.5 mb, which is quite small as
compared to the corresponding value of $\sigma_{-2n}^{exp}$ \cite{aum}. This 
difference too can be understood from the fact that the  
cross sections corresponding to the nuclear elastic and inelastic breakup
modes are quite large \cite{gar00b,ers00}. 

The  pure Coulomb two-neutron removal
cross sections, calculated within the dineutron model (open squares), 
are even larger than $\sigma_{-2n}^{exp}$, 
and have a wrong type of the beam energy dependence. 
Thus, the dineutron model is not suitable for  
describing the data for the breakup of $^6$He.  

\subsection{$\alpha$-particle momentum distributions}

\begin{figure}[ht]
\begin{center}
\mbox{\epsfig{file=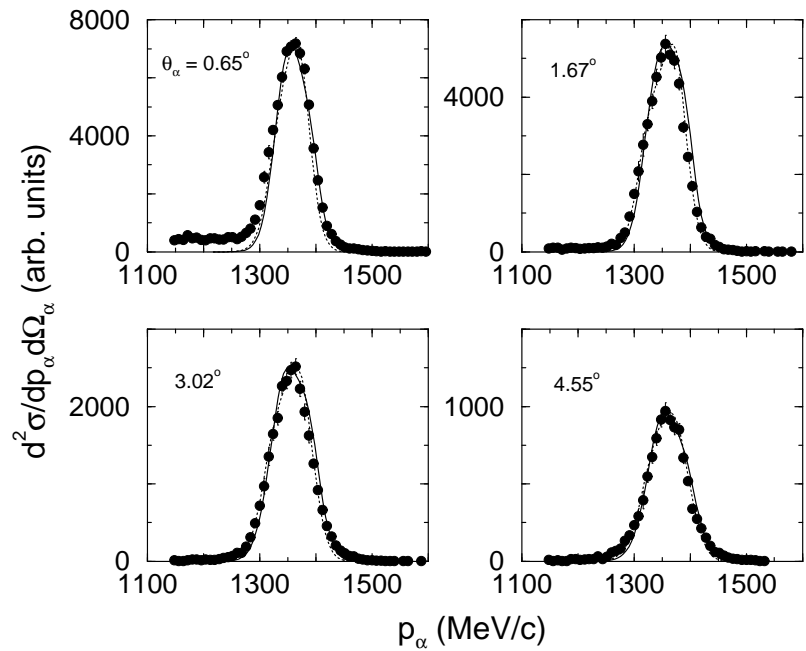,width=.75\textwidth}}
\end{center}
\caption{ Comparison of the calculated and measured total  
momentum distributions of the alpha particle in the breakup of 
$^6$He on a Au target at 64 MeV/nucleon at four forward angles.   
The results of the FRDWBA and AD models are shown
by the solid and dotted lines, respectively.     
The data are taken from \protect\cite{bush}.}   
\label{fig:fig4}
\end{figure}

In Fig. 4, we compare the calculated and measured \cite{bush} 
total momentum distributions of the $\alpha$-particle emitted in the  
breakup of $^6$He on the Au target at the beam energy of 64 MeV/nucleon. The 
$\alpha$-particle was detected at four average angles of 0.65$^{\circ}$,
1.67$^\circ$, 3.02$^\circ$ and 4.55$^\circ$ \cite{bush}. In this 
experiment, the  coincident $\gamma$-ray technique was used to identify
events in which the target nucleus was simultaneously excited. The 
elastic breakup events were tagged with $\gamma$-ray
multiplicity $M_{\gamma}$ = 0. Since the data are given in arbitrary units,
we have normalized our calculations to the peak of the experimental 
distributions in all four cases. It is clear that shapes of the momentum
distributions are well reproduced by our calculations. The full width at
half maximum (FWHM) of the momentum distributions 
changes systematically from 76 MeV/c to 98 MeV/c with increasing angle. 
This could indicate that the effects related to the reaction dynamics
({$\it e.g.$} distortion of the wave functions describing the fragment-target
relative motion) lead to the widening of
the total momentum distributions as a function of the detection angle.

It should be remarked
here that the calculations \cite{zhuk91}, where the total momentum
distributions are obtained from the momentum space wave functions of the
ground state of the projectile disregarding the reaction dynamics
altogether, are unlikely to describe this feature of the experimental data.  
At the same time, if the detection angle increases further the nuclear breakup 
effects will become more important and pure Coulomb breakup may
also become inadequate to describe the FWHM of the experimental total 
momentum distributions. Therefore, it would be interesting to span a
larger range of detection angles in future experiments.  

\begin{figure}[ht]
\begin{center}
\mbox{\epsfig{file=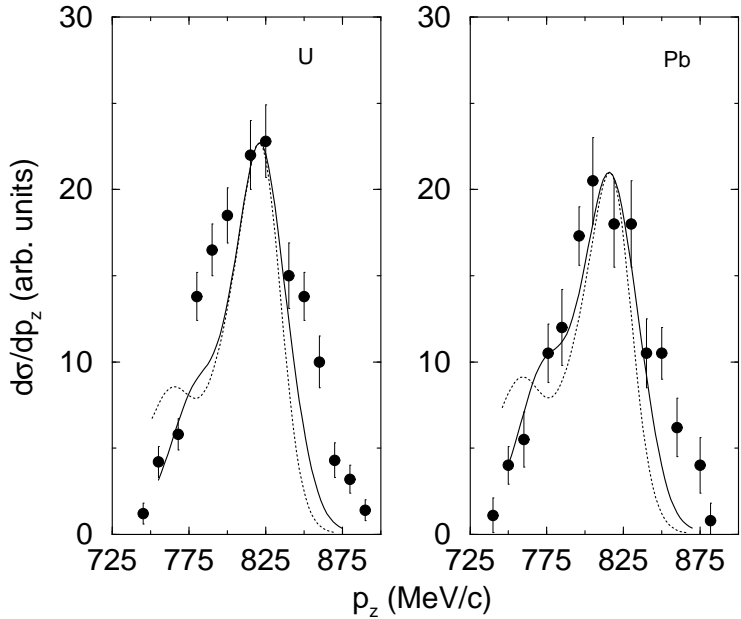,width=.70\textwidth}}
\end{center}
\caption{
Parallel momentum distribution of the alpha particle emitted in
the Coulomb breakup of $^6$He on U and Pb targets at the beam energy of
24 MeV/nucleon.  The FRDWBA and AD model results are shown
by the solid and dotted lines respectively.     
The data are taken from \protect\cite{aar}.}   
\label{fig:fig5}
\end{figure}

In Fig. 5, we present the comparison of our calculations with the data 
for the parallel momentum (p$_z$) distribution of the $\alpha$-particle
emitted in the breakup reaction of $^6$He on U and Pb targets at the
beam energy of 24 MeV/nucleon. These measurements have 
recently been performed at the Michigan State University \cite{aar}.
The calculated distributions have been obtained by integrating Eq. (20)
over the neutron angles up to 5.7$^\circ$ (which is the minimum opening
angle of the neutron detector in this experiment \cite{aar}).
The calculated results are normalized to the peak of the data.
We note that although both FRDWBA
and AD models are able to reproduce the FWHM of the experimental p$_z$
distributions ($\sim$ 76 MeV/c) reasonably well, the former is in slightly
better agreement with the data.
\subsection{Neutron-neutron and core-neutron relative energy spectra}
\begin{figure}[ht]
\begin{center}
\mbox{\epsfig{file=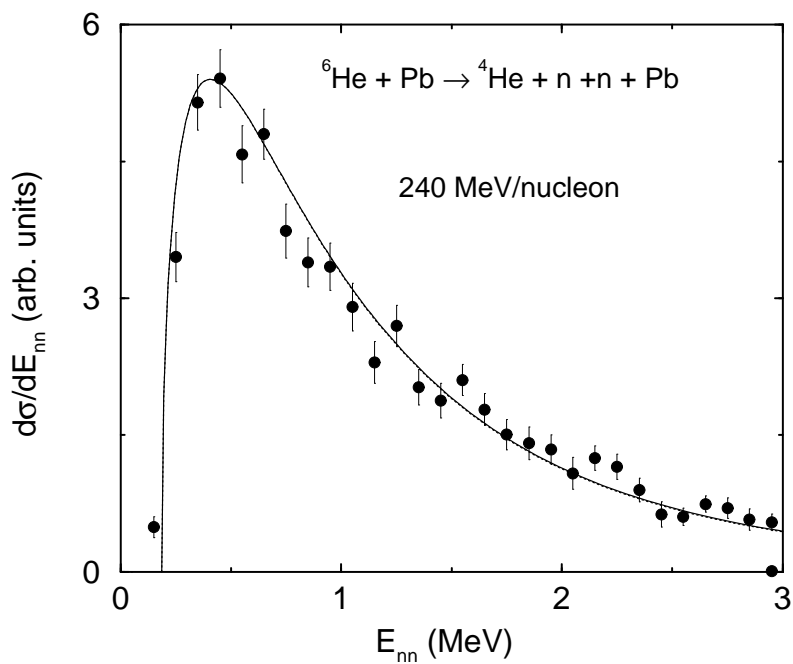,width=.75\textwidth}}
\end{center}
\caption{Calculated
neutron-neutron relative energy spectra in the Coulomb breakup of 
$^6$He on the Pb target at the beam energy of 240 MeV/nucleon. 
The FRDWBA and  adiabatic model results are shown
by the solid and dotted lines respectively.     
The data are taken from \protect\cite{aum}.}   
\label{fig:fig6}
\end{figure}

In Fig. 6, we show the comparison of our calculations with the 
data \cite{aum} for the neutron-neutron relative energy (E$_{nn}$) spectrum
obtained in the breakup of $^6$He on a Pb target at the beam energy of
240 MeV/nucleon. The calculated spectrum has been obtained by transforming
the cross section shown in Eq. (20) to the frame of relative and c.m. 
motion of the fragments (see Appendix B), and integrating the resultant 
over the unobserved variables. The integrations over the theta
and phi angles of $n_1$-$n_2$ and $\alpha$-($n_1$-$n_2$) systems were
carried out in the range of 0$^\circ$-180$^\circ$ and
0$^\circ$-360$^\circ$ respectively, while
those  over the energy E$_{\alpha-(n_1-n_2)}$ are done in the range
of 0-10 MeV. However, the integration over the theta angle of the
$^6$He-target system ($\theta_{at}$ in Eq. (B.12))
is carried out up to the grazing angle (1.1$^\circ$) only, as we are interested
only in the pure Coulomb contributions to the cross sections.  
The solid (dotted) line shows the corresponding FRDWBA (AD) model results.
Both calculations are normalized to the peak of the experimental data
which is at E$_{nn}\sim$ 0.45 MeV. Both the models lead to almost similar
results which incidently are in good agreement with the shape of the
experimental data. Of course, inclusion of the nuclear
and Coulomb-nuclear interference terms can affect both 
the absolute magnitude as well as shape of this distribution. Therefore,
it would be interesting to have the data with absolute values on the one
hand, and the calculations including both Coulomb and nuclear breakup 
effects on the other.  

The shape of the neutron-neutron
energy spectrum reported in Ref. \cite{zhuk00} is not in agreement
with that of the experimental data. This is due to the fact that these
authors have considered a pure sudden approximation, and have ignored 
the final state interaction effects. This underlines the
importance of the reaction dynamics in the breakup reactions even for
beam energies around  240 MeV.  
 
\begin{figure}[ht]
\begin{center}
\mbox{\epsfig{file=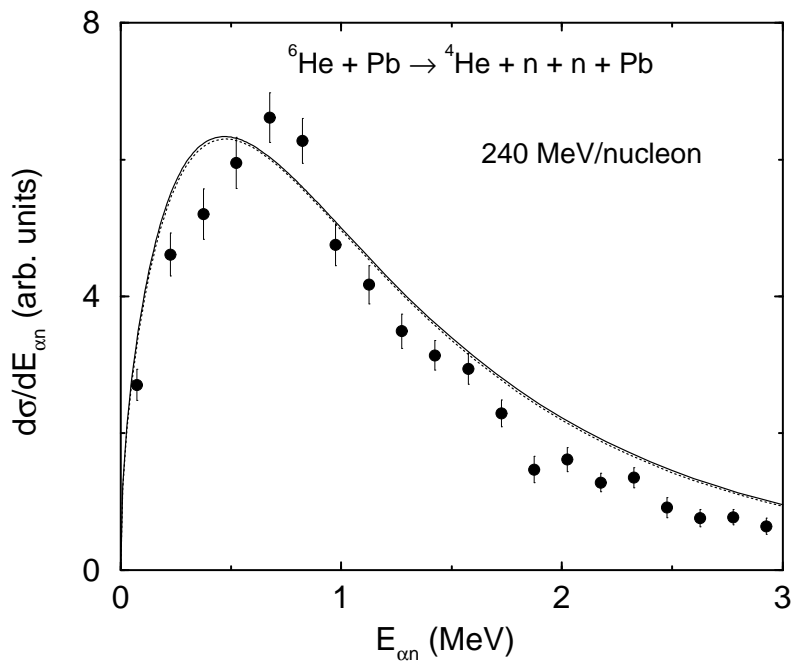,width=.75\textwidth}}
\end{center}
\caption{Calculated
$n$-$\alpha$ relative energy spectra in the Coulomb breakup of 
$^6$He on the Pb target at the beam energy of 240 MeV/nucleon. 
The FRDWBA and AD model results are shown
by the solid and dotted lines respectively.      
The data are taken from \protect\cite{aum}.} 
\label{fig:fig7}
\end{figure}

In Fig. 7, we have compared the calculated $n$-$^4$He relative energy 
(E$_{\alpha n}$) spectrum  with the corresponding experimental data \cite{aum}.
The peaks of the theoretical calculations are normalized to the maximum of
the experimental data. The theoretical cross sections have been
obtained by integrating Eq. (B.14) of Appendix B over the 
unobserved variables. The range of the angular and the energy integrations
of various variables were the same as that described above in
connection with Fig. 6. We see that except for the peak position, the
theoretical calculations reproduce the data reasonably well.
The difference in the peak positions of the experimental and theoretical 
spectra may perhaps be attributed to the fact that the position of the
$^5$He resonance is not properly reproduced within the three-body model
used in getting the ground state wave function of $^6$He
employed in our calculations. We plan to check this point in a future
study by using other $^6$He wave functions which may be better in
this respect.
\subsection{Excitation energy spectra}
\begin{figure}[ht]
\begin{center}
\mbox{\epsfig{file=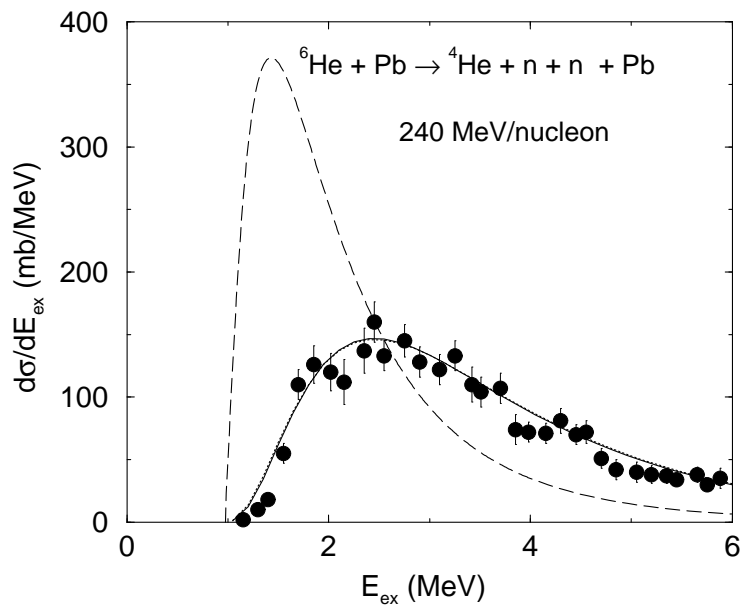,width=.70\textwidth}}
\end{center}
\caption{Calculated
excitation energy spectra in the Coulomb breakup of 
$^6$He on the Pb target at the beam energy of 240 MeV/nucleon. 
The FRDWBA and AD model results are shown
by he solid and dotted lines respectively.     
The data are taken from \protect\cite{aum}. The dashed line shows the
result of the dineutron model.}   
\label{fig:fig8}
\end{figure}

In Fig. 8, we show the pure Coulomb contribution to the 
excitation energy (E$_{\rm ex}$) spectrum in the breakup of $^6$He on 
a Pb target at the  beam energy of 240 MeV/nucleon  \cite{aum}. 
Excitation energy is defined as the sum of the $\alpha$-($n_1$-$n_2$)
and  $n_1$-$n_2$ relative energies and the $Q$-value of the reaction.
The calculated cross sections are normalized to the peak of the data as the
elastic Coulomb breakup contribution constitutes only a part of the 
excitation energy spectrum \cite{ers00}. We note that both 
FRDWBA and AD models give identical results and are able to reproduce
the shape of the spectrum well. We have also shown in this figure
the results obtained with the dineutron model, which is clearly inadequate
to describe the data. It should, however, be noted that
the calculated cross sections have not been convoluted with the
detector response matrix as suggested in \cite{aum}. As is  
shown in \cite{aum}, this folding reduces the 
cross sections by about 20 $\%$ near the peak region, and by less that 
5$\%$ beyond it. Therefore, the agreement with the data seen in 
Fig. 8 should be seen in this context. 

\section{Summary and Conclusions}

In this paper we have calculated the elastic Coulomb breakup of the
Borromean nucleus $^6$He within a four--body post form distorted wave Born
approximation approach. The finite range effects have been included
by using an effective local momentum approximation on the Coulomb
distorted wave function of the charged heavy core in the outgoing channel.
This method allows the use of realistic correlated three-body wave 
functions for the projectile. Thus the special features 
associated with the structure of the two-neutron halo systems can be
taken into account in the reaction calculations. In our method, the breakup
amplitude is expressed as a product of the factors describing separately the
projectile structure and the dynamics of the reaction. This feature of
our theory is common with that of the adiabatic model of the Coulomb breakup
reactions. Although unlike DWBA, this model does not use the weak coupling
approximation to describe the center of mass motion of the fragments with
respect to the target, yet it leads to results which agree with those 
of the FRDWBA within approximately 20$\%$. The existing data are unable 
to highlight the significance of this small difference between the two 
calculations. Measurements done below the grazing angle may be 
helpful in this respect. 

We have presented calculations for the energy,
total and parallel momentum distributions of the $\alpha$-particle
as well as for the 
neutron-neutron and $\alpha$-neutron relative energy spectra.  
In all the cases, our calculations are able to reproduce the
shapes of the experimental spectra reasonably well. However,
the absolute magnitudes of the data (wherever available) were 
underpredicted by both finite range DWBA and adiabatic model of the 
Coulomb breakup. This suggests that nuclear breakup 
contributions are substantial in the kinematical regimes covered
by these data. Furthermore, although the pure Coulomb calculations
are able to reproduce the shapes of the experimental data on the 
fragment-fragment correlations, inclusion of nuclear breakup effects
are necessary for a more comprehensive description of the 
data.  Our calculations rule out the point-like dineutron-core type of
models for describing the breakup of Borromean systems. The 
two-body fragment-fragment correlation data are beyond the scope
of such models.     

The authors are thankful to Ian Thompson of the University of Surrey 
for providing them the three-body wave function of $^6$He used 
in the calculations reported in this paper and to Tom Aumann and Hans
Emling of GSI, Darmstadt for making them available in tabular form
the experimental data reported in Figs. 6,7, and 8. One of us (RS)
wishes to thank Pawel Danielewicz for the kind hospitality of the
theory group of the National Superconducting Cyclotron Laboratory
of the Michigan State University where this paper was finalized.
He also wishes to thank Aaron Galonsky for several useful discussions
regarding the experimental data reported in Fig.5 and for 
going through the manuscript.
       
\appendix
 
\section{ Validity of the effective local momentum approximation}
 
\begin{figure}[ht]
\begin{center}
\mbox{\epsfig{file=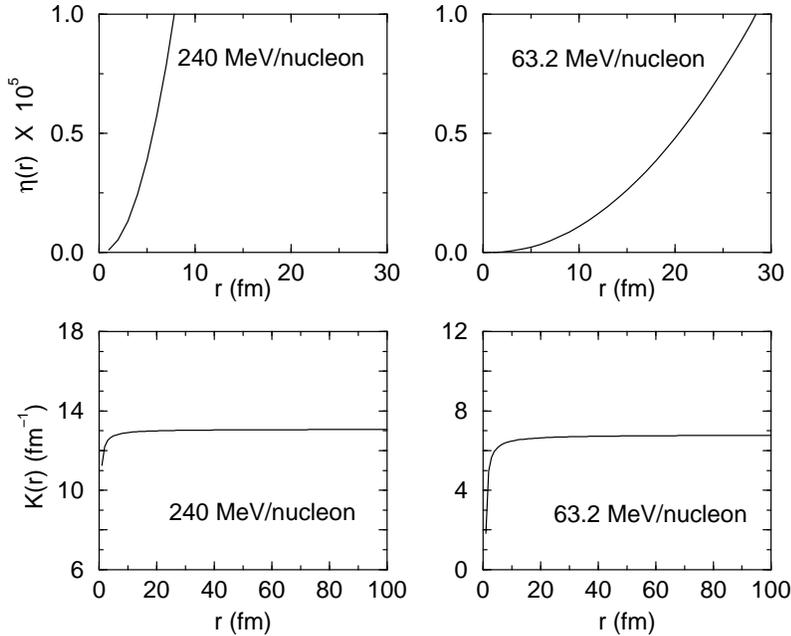,width=.75\textwidth}}
\end{center}
\caption{Variation of $\eta (r)$ (upper half) and ${ K}(r)$
(lower half) with $r$ for the reaction
$^6$He + $^{197}$Au $\rightarrow$ $^4$He + $n$ + $n$ + $^{197}$Au
at the beam energies of 240 MeV/nucleon (left side) and 63.2 MeV/nucleon 
(right side).
}
\label{fig:figo}
\end{figure}
The derivation of the finite range amplitude [Eq. (13)] uses an 
effective local momentum approximation (ELMA). This allows us to write
the distorted wave function of argument $\rb$ as a product
of the plane wave of argument ${\bf R}$ and distorted wave of 
argument $\ri$. This simplifies the evaluation of the six-dimensional
integrals by reducing them as a product of two three-dimensional
integrals of independent arguments.

As discussed in \cite{shyam85}, a condition for the validity of this 
approximation is that the quantity 
\begin{eqnarray}
\eta(r) = {1\over 2}{ K}(r)|d{ K}(r)/dr|^{-1}
\end{eqnarray}
evaluated at a representative distance $R_d$ should  
be much larger than the rms radius of the projectile. 
To check this, we show in Fig. A.1, 
$\eta(r)$ (upper half) and ${ K}(r)$ (the magnitude of the local
momentum) (lower half) as a function of $r$, for the breakup reactions
$^6$He + $^{197}$Au $\rightarrow$ $^4$He + $n$ + $n$ + $^{197}$Au
at the beam energies of 240 MeV/nucleon (left side) and 63.2 MeV/nucleon 
(right side).  We see that for $r$ $>$ 8 fm,
$\eta(r)$ is several orders of magnitude larger than rms radii of the
halo in both the cases. Therefore, the above condition is well satisfied.  

>From the lower half of Fig. A.1, we note that the value of ${ K}(r)$
remains constant for distances larger than 10 fm for both the
beam energies. Due to the peripheral
nature of the breakup reaction, this region contributes maximum to the 
cross section. Therefore, our choice of a constant  magnitude for
the local momentum evaluated at 10 fm is well justified and is independent
of the beam energy. In fact, we noted 
that as $R_d$ is increased from 5 to 10 fm the calculated cross sections
vary by at the most 10 $\%$, and with a further increase the variation
is less than 1 $\%$, in all the cases considered in this paper. 
\begin{figure}[ht]
\begin{center}
\mbox{\epsfig{file=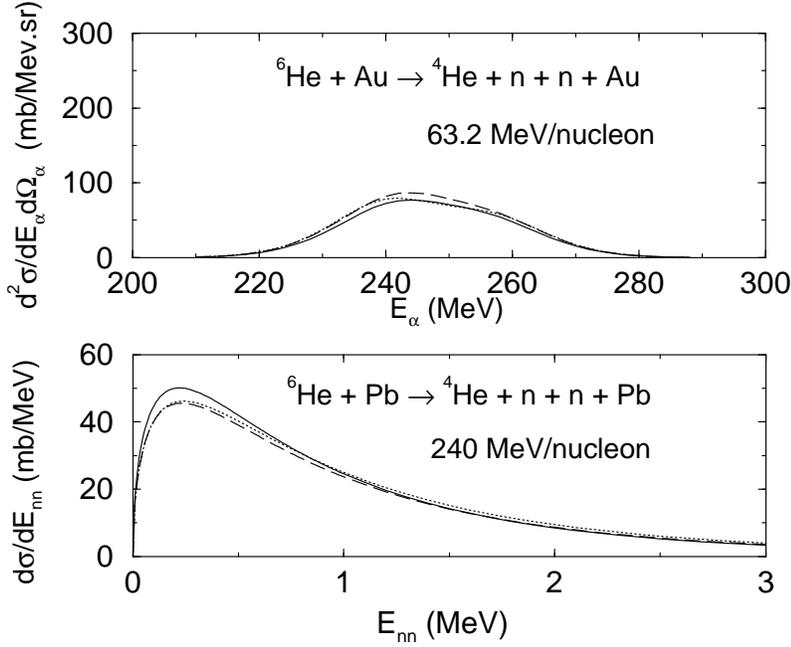,width=.75\textwidth}}
\end{center}
\caption{The energy distributions of the alpha particle  
in the reaction $^6$He + $^{197}$Au $\rightarrow$ $^4$He + $n$ + $n$ + 
$^{197}$Au at the beam energy of 63.2 MeV/nucleon (upper part),
and the neutron-neutron energy correlation in the reaction  
$^6$He + $^{208}$Pb $\rightarrow$ $^4$He + $n$ + $n$ + $^{208}$Pb
at the beam energy of 240 MeV/nucleon (lower part),  
for choices $d_1$ (solid line), $d_2$ (dashed line) and $d_3$ (dotted line)
for the  direction of the local momentum, as discussed in the text. 
} 
\label{fig:figp}
\end{figure}

In the application of the ELMA in the description 
of the heavy ion induced transfer reactions \cite{braun74b}
and the breakup reaction of the one-neutron halo nuclei \cite{raja00},  
the calculated cross sections were found to be almost unaffected by the
choice of the direction of the local momentum. To study the sensitivity
of our FRDWBA results on the direction of ${ {\bf K}}$ for reactions
under study in this paper, we have performed calculations for
three cases where we take the angles of the local momentum
($d_1$) parallel to those of ${\bf k}_b$, ($d_2$) parallel to the direction
corresponding to the half of the angles of ${\bf k}_b$ and ($d_3$) parallel
to the beam direction (zero angles).
In the upper part of Fig. A.2, we show the energy distribution
of the alpha particle for the same reaction as described in Fig. 2.  
The results obtained with cases $d_1$, $d_2$ and $d_3$ are shown
by solid, dashed and dotted lines respectively. We note that energy
distributions calculated with these choices are almost identical. 
The results for the neutron-neutron energy correlations  
for the same reaction as in Fig. 6, are shown in the lower
part of this figure for the choices ($d_1$), ($d_2$) and ($d_3$) for
the direction of the local momentum. In this case too we observe almost
no variation in the calculated cross sections. Similarly, we have noted
no dependence of the calculated widths of the  parallel momentum
distributions of heavy fragments on the direction of ${ {\bf K}}$, in all
the reactions investigated in this paper. Therefore, the dependence of
various cross sections for the reactions studied in this paper, on the
direction of the local momentum is either very minor or almost negligible.
The measurements done so far are not able to distinguish the small
differences that we see here in some cases. Hence, we have 
performed all our calculations in this paper by using the direction 
of the momentum ${\bf K}$ to be the same as that of ${\bf k}_b$.

\section{ The four -- body phase space factor for a given Q value} 

The phase space in which we fix the total energy $E_{tot}$
in the final state at the value $E$ is given by
\begin{eqnarray}
 d\sigma = h^{-9}\delta(E_{tot} - E)d{\bf p}_bd{\bf p}_{n_1}d{\bf p}_{n_2}
\end{eqnarray}
The total energy is given by
\begin{eqnarray}
E_{tot} &=& E_a + Q \\
        &=& E_t + E_b + E_{n_1} + E_{n_2} \\ 
&=& E_b + E_{n_1} + {p^2_{n_2} \over {2m_{n_2}}} +
   {p^2_{t} \over {2m_{t}}} \\
&=&  E_b + E_{n_1} + {p^2_{n_2} \over {2m_{n_2}}} +
 {1\over{2m_t}}[({\bf p}_{tot}-{\bf p}_{b}-{\bf p}_{n_1})^2 
 + p^2_{n_2} \\ \nonumber
 &-& 2{\bf p}_{n_2}.({\bf p}_{tot}-{\bf p}_{b}-{\bf p}_{n_1})],           
\end{eqnarray}
where $E_a$ is the projectile energy and $Q$ is the 
Q-value of the reaction ($<$0). ${\bf p_{tot}}$ is the total
momentum of the projectile in the incident channel, which by 
momentum conservation, is given by 
\begin{eqnarray}
{\bf p}_{tot} = {\bf p}_{t} + {\bf p}_{b} + {\bf p}_{n_1} + {\bf p}_{n_2}.
\end{eqnarray}
Using $d{\bf p} = m.p.dE.d\Omega$, we have
\begin{eqnarray}
{{d\sigma}\over {dE_b d\Omega_b dE_{n_1} d\Omega_{n_1} 
dE_{n_2} d\Omega_{n_2}}} &=& m_b p_b m_{n_1}p_{n_1}m_{n_2}p_{n_2}
 {{d\sigma}\over {d{\bf p}_{b}d{\bf p}_{n_1}d{\bf p}_{n_2}}} \\
 &=& h^{-9}\delta(E_{tot} - E)m_b p_b m_{n_1}p_{n_1}m_{n_2}p_{n_2} 
\end{eqnarray}
Therefore
\begin{eqnarray}
\rho(E_b,\Omega_b,E_{n_1},\Omega_{n_1},\Omega_{n_2}) &=&
{{d\sigma}\over {dE_b d\Omega_b dE_{n_1} d\Omega_{n_1} d\Omega_{n_2}}}
\nonumber \\
&=& \int {{d\sigma}\over {dE_b d\Omega_b dE_{n_1}d\Omega_{n_1} 
dE_{n_2} d\Omega_{n_2}}} dE_{n_2} \nonumber\\ 
 &=&h^{-9}m_bm_{n_1}m_{n_2}
 \int p_bp_{n_1}p_{n_2}\delta(E_{tot}-E)dE_{n_2}
\nonumber \\ 
&=& h^{-9}m_bm_{n_1}m_{n_2}\Big[{p_bp_{n_1}p_{n_2}\over
{\partial E_{tot}/\partial E_{n_2}}}\Big]_{E_{tot} = E}
\end{eqnarray}
Now from Eq.(B.5)
\begin{eqnarray}
{\partial E_{tot}\over \partial E_{n_2}} = 
{\partial E_{tot}\over \partial p_{n_2}}{dp_{n_2}\over dE_{n_2}} = {1\over m_t}[m_{n_2} +
m_t -m_{n_2}{{\bf p}_{n_2}.({\bf p}_{tot} -{\bf p}_b - 
{\bf p}_{n_1})\over p_{n_2}^2}]
\end{eqnarray}
Therefore
\begin{eqnarray}
\rho(E_b,\Omega_b,E_{n_1},\Omega_{n_1},\Omega_{n_2}) = 
{h^{-9}m_tm_bm_{n_1}m_{n_2}p_bp_{n_1}p_{n_2}\over m_{n_2} +
m_t -m_{n_2}{{\bf p}_{n_2}.({\bf p}_{tot} -{\bf p}_b - {\bf p}_{n_1})\over p_{n_2}^2}}.
\end{eqnarray}

It can also be shown that
\begin{eqnarray}
{{d^5 \sigma}\over {dE_{n_1n_2} d\Omega_{n_1n_2} dE_{b-({n_1n_2})}
 d\Omega_{b-({n_1n_2})}d\Omega_{at} }} &=& J_1 
\nonumber \\ 
&\times & {{d^5\sigma}\over {dE_b d\Omega_b dE_{n_1}
d\Omega_{n_1} d\Omega_{n_2}}},
\end{eqnarray}
where
\begin{eqnarray}
J_1&=&{{{m_{n_1n_2}p_{n_1n_2}m_{b-({n_1n_2})}p_{\alpha-({n_1n_2})}
m_{at}p_{at}}\over {h^{9}\rho(E_b,\Omega_b,E_{n_1},\Omega_{n_1},\Omega_{n_2})}}}
\end{eqnarray}
and
\begin{eqnarray}
{{d^5 \sigma}\over {dE_{bn_1} d\Omega_{bn_1} dE_{n_2-({bn_1})}
 d\Omega_{n_2-({bn_1})}d\Omega_{at} }} &=& J_2 
\nonumber \\ 
&\times &{{d^5\sigma}\over {dE_b d\Omega_b dE_{n_1}
d\Omega_{n_1} d\Omega_{n_2}}},
\end{eqnarray}
where
\begin{eqnarray}
J_2&=&{{{m_{bn_1}p_{bn_1}m_{n_2-({bn_1})}p_{n_2-({bn_1})}
m_{at}p_{at}}\over {h^{9}\rho(E_b,\Omega_b,E_{n_1},\Omega_{n_1},
\Omega_{n_2})}}}.
\end{eqnarray}

\end{document}